\documentclass[fleqn,10pt]{wlscirep}
\usepackage[utf8]{inputenc}
\usepackage[T1]{fontenc}
\usepackage{url}
\usepackage{color}
\def\beq{\begin{equation}}
\def\eeq{\end{equation}}
\title{Viscous wave models to retrieve the thickness of grease-pancake sea ice using SAR image wave spectra}
\title{SAR image wave spectra to retrieve the thickness of grease-pancake sea ice using viscous wave models}

\author[1*]{Giacomo De Carolis}
\author[2]{Piero Olla}
\author[1]{Francesca De Santi}
\affil[1]{National Research Council of Italy, Institute for Electromagnetic Sensing of the Environment, Milan, 20133, Italy}
\affil[2]{National Research Council of Italy, Institute of Atmospheric Science and Climate, Cagliari, 09042, Italy}

\affil[*]{giacomo.decarolis@cnr.it}

\begin{abstract}
Young sea ice composed of grease and pancake ice (GPI), as well as thin floes, considered to be the most common form of sea ice fringing Antarctica, is now  becoming the “new normal” also in the Arctic. Investigations to determine how an increase in GPI is affecting the climate in the far north and globally, require specific tools to monitor the GPI’s thickness distribution.
Directional wave spectra from satellite SAR imagery are used to determine the change in wave dispersion as a wave train enters GPI fields. The ice cover thickness is then estimated by fitting the 
dispersion data with two models of wave propagation in ice cover ocean: the Keller's model
and the close-packing (CP) model.
For both models, an empirical constitutive equation for GPI viscosity as a function of the ice thickness is derived and discussed.
Examples of GPI thickness retrievals are shown for a Sentinel-1 C band SAR image taken in the Beaufort Sea on 1 November 2015, and three CosmoSkyMed X band SAR images taken in the Weddell Sea on March 2019. The estimated GPI thicknesses are consistent with concurrent SMOS measurements and the available local samplings.
\end{abstract}
\begin{document}

\flushbottom
\maketitle
%
%
\thispagestyle{empty}

\section*{Introduction}

Climate change has led 
 in the last decade to a dramatic reduction
of the extent and the thickness
of sea ice in the Arctic during the summer season. As a result, 
the marginal ice zone (MIZ), i.e. the dynamic transition region from the dense inner pack ice zone to open ocean, resulted even more directly exposed to the wind and wave action\cite{thomson2014swell}.
Grease and pancakes ice (GPI) are thus becoming the dominant ice types in the Arctic seas during the freezing season\cite{thomson2018overview}. In consideration of the high production of GPI in the southern oceans \cite{doble2003pancake}, GPI can then be considered one of the most important sea ice types in the polar oceans. Investigations to determine how an increase in GPI may affect the 
climate, possibly leading to local warming in  the far north, are ongoing.
These studies and their implementation in climate models clearly require some effective tools to monitor the GPI’s properties, especially as regards its thickness.

As the extent of GPI fields is controlled by the presence of waves that usually come from the open ocean, the relationship between wave attenuation rates and GPI rheology is a key factor to understand and model the evolution of the ice cover. Given the large scale separation between pancakes and the typical wavelength, a good approximation to describe the interaction with ocean waves is to represent GPI as a continuum \cite{peters1950effect, keller1998gravity}. Validation and calibration of any GPI wave model are tasks that cannot be entrusted to in situ activities, due to the vast extent  of the GPI fields, their intrinsic dynamic nature, and the harsh environment where they develop. 
A systematic and extensive characterization of the spatio--temporal distribution of GPI can however be carried on through remote sensing by exploiting the capability of SAR imaging techniques to measure the full ocean wave directional spectrum \cite{hasselmann1991nonlinear}. By tracking the SAR observed ocean wave spectrum (or its peak, as done in pioneer studies such 
as\cite{wadhams1991waves}), from open sea throughout the GPI cover, it is indeed possible to measure wave dispersion and attenuation.
A dedicated GPI wave model can then be used to relate the modifications in the wave propagation with
GPI properties such as its thickness and mechanical properties.

The literature on the application of SAR spectral inversion techniques to GPI thickness retrieval
dates back at least to 1997.
In earlier papers \cite{ wadhams1997wave,  wadhams2002use, wadhams1999mapping, de2001retrieval}, the GPI thickness was estimated by a adopting a mass-loading scheme,  in which
GPI is represented as a continuum of non-interacting point-like mass loads on the sea surface \cite{peters1950effect}. The success of the approach, however, was limited by the excessively
high values of the thickness resulting from SAR inversion.

Later papers \cite{decarolis2003sar, wadhams2004sar, wadhams2018pancake} abandoned the mass-loading approach for the more realistic representation of  GPI  proposed by Keller \cite{keller1998gravity}. The Keller's model pictures the GPI layer as a viscous fluid of given thickness $h$ and effective viscosity $\nu$ floating over an infinitely deep, inviscid water column. The GPI effective viscosity is an unknown parameter, with experimental evidence suggesting strong variability with respect to the composition of the sea ice matrix.
As an example, wave buoy attenuation data collected in the Weddell Sea over GPI fields with pancakes thicknesses up to 50 cm,  compared with the Keller's models, revealed a variability of GPI effective viscosity values spanning the range $10^{-1}-10^{3}  \rm{m^2/s}$\cite{doble2015relating},  
without a clear relationship with the GPI layer thickness. 

Other viscous layer models have been proposed to describe wave propagation in ice/water systems. 
We can mention the two–-layer viscous models (TLV), which assumes the water underneath the ice
cover to have a finite viscosity, possibly due to turbulent effects
\cite{de2002dispersion}, and the viscoelastic model, which treats the ice layer as a viscoelastic
medium \cite{wang2010gravity}. All these generalizations of the Keller's model, however,
seem unable to significantly reduce the variability of the effective ice viscosity \cite{cheng2017calibrating,de2018ocean}.

We argue that at least part of the variability of the effective viscosity required by the Keller's 
model to describe the properties of GPI, may stem from attempting to treat the ice layer as 
a homogeneous medium.
The close-packing (CP) model\cite{de2017effect} makes a first attempt to explicitly take into
account the inhomogeneity of GPI, by
assuming that pancakes are confined to a fictitious layer of infinitesimal thickness, lying on top of
the grease ice layer and modifying the wave stress at the upper surface. 
The CP model was tested on field data from the Arctic and the Antarctic, along with Keller's and TLV models\cite{de2018ocean}. 
Fitting of the field data was possible with all the models considered, but
only the CP model produced GPI effective viscosities in a range consistent with that of grease ice in laboratory experiments ($\approx 2.5-3 \times 10^{-2} {\rm m^2/s}$)  \cite{newyear1999comparison}.

The SAR inversion procedure consists of a three-step process: (i) simulation of the ocean wave spectrum in sea ice, with the selected GPI wave propagation model and given values of $\nu$ and $h$, (ii) transformation of the wave spectrum to a SAR image spectrum \cite{hasselmann1991nonlinear}, (iii) comparison between the modeled and observed SAR spectra by means of an appropriate cost function.

In former SAR inversion schemes, the couple $(h,\nu)$ minimizing the distance---parameterized
by the cost function---between simulated and observed SAR image spectrum, was then 
assigned to the GPI portion traveled by the waves.
However, we have observed that the intrinsic structure of GPI viscous wave models does not 
allow to identify an absolute minimum of the cost function. Rather, deep valleys in the parameters' space 
can be observed, which leads to important consequences on the procedure of ice thickness retrieval 
\cite{de2018ocean}. The valleys follow power-law curves $\nu \propto h^\alpha$, 
where $\alpha$ is characteristic of the specific GPI wave model considered, and result
in an underdetermined minimization problem. In order to infer a unique minimizing couple ($h,\nu$),  a  relationship between effective ice viscosity and thickness has to be empirically formulated through a calibration process.

We perform such calibration thanks to a study conducted in the Odden Ice Tongue, Greenland Sea, that developed in 1996/1997 with massive presence of GPI, extensively documented by SAR imagery,  and for which GPI thickness maps obtained from a specifically developed model were available\cite{pedersen2004sea}. Both in the case of the Keller's and the CP model, calibration produces an effective viscosity which is a monotonically increasing function of the ice thickness.

We have applied the calibrated SAR inversion schemes to two case studies in the Arctic and in Antarctica, using SAR images gathered by Sentinel-1A and COSMO-SkyMed, respectively.

\section*{Results}
The analysis of a set of wave buoy data gathered in the advancing MIZ of the Weddell Sea covered by GPI clearly revealed that the GPI wave attenuation rates scale with the “equivalent solid ice 
thickness” \cite{doble2015relating}
\begin{equation}
    h=C_{gr}h_{gr}+C_ph_p,
\label{heq}
\end{equation}
where $C$ and $h$ are respectively 
the concentration and thickness of grease ($g$r) and pancake ($p$) ice. 
The relation is physically sound, as it states a proportionality between wave decay and ice volume 
per unit area of the region of the sea traveled by the waves. 
We point out that the effective ice thickness is the quantity most relevant to the overall changes 
in ice volume, and is used in numerical dynamic-thermodynamic sea ice models 
\cite{holland2014modeled}. Hereafter we will refer to the GPI thickness as the effective ice 
thickness of the GPI layer.

The analysis also reveals that 
an important component of the GPI's viscosity variability can be explained through 
$h$ dependence of $\nu$, which allows one
to refine SAR inversion schemes such as the ones by De Carolis 
(2001 and 2003) \cite{de2001retrieval, decarolis2003sar}.


\subsection*{SAR calibration procedure of GPI thickness vs viscosity}
The predictions on the effective viscosity of sea ice provided by the TLV and the viscoeleastic models
have been carried out elsewhere and showed not to differ significantly from the ones by the Keller's model 
\cite{de2018ocean,cheng2017calibrating}.
We thus carry out the calibration procedure limited to the Keller's and the CP model.

\begin{figure}[htbp]
\centering
\includegraphics[width=\linewidth]{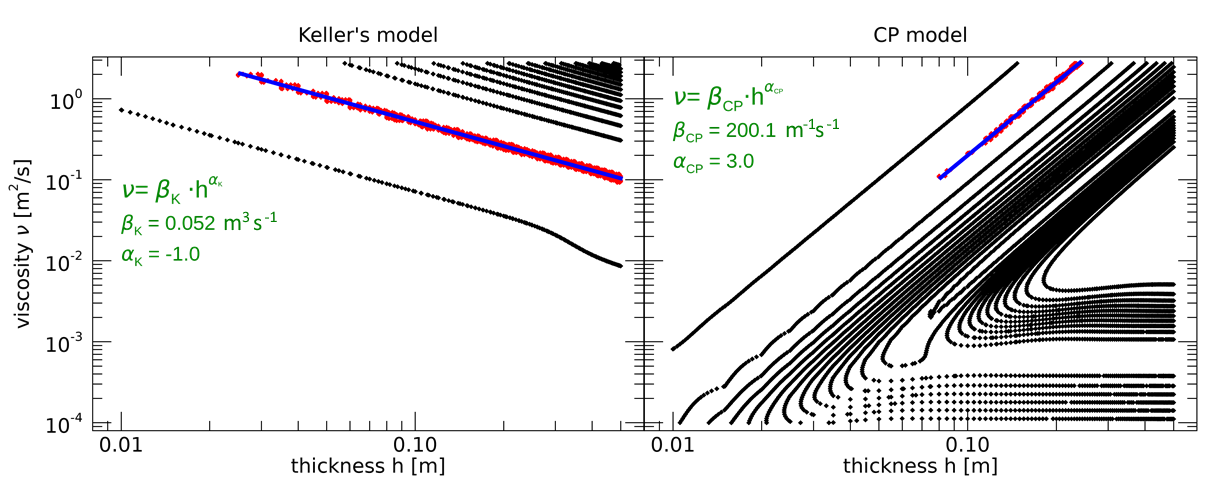}
\caption{Exemplary contour plots of the cost functions obtained from SAR inversion in GPI field for the Keller's model (left panel) and the CP model (right panel). Red points mark those locations $(h,\nu)$ where the cost function values are within 1\% of the corresponding absolute minimum. This example corresponds to the window number 13 selected on the ERS2 SAR image of the 9 March 1997 and is representative for any cases considered in this paper.}
\label{fig:cf}
\end{figure}

The calibration proceeds through minimization of a cost function, defined in Eq. (\ref{Psi}), 
which gives the $L^2$ distance of the observed and the simulated SAR image spectra.
As discussed in the Methods Section, the cost function minima define power-law
curves in the ($h,\nu$) plane
\begin{equation}
\nu=\beta h^\alpha,			
\label{hnu}	
\end{equation}
where $\alpha$ is specific of the GPI model utilized:
\beq
\alpha_{CP}=3,
\qquad
\alpha_K=-1,
\label{alphas}
\eeq
respectively, for the CP and the Keller's model\cite{de2018ocean},
while $\beta$ depends on both the model and the actual SAR image considered.
Examples of the cost functions pictured as contour lines are shown Figure \ref{fig:cf}.

%
%
We note that  although Eq. (\ref{hnu}) can be analytically proved only for small $\hat h$, 
the contour lines in Fig. \ref{fig:cf} are obtained by simulating the SAR wave spectrum with 
the full wave dispersion relation.
As this trend is common to all the examined cases, we assume that Eq. \ref{hnu} always holds.

The parameter $\beta$ is strongly dependent on the specific SAR image portion examined. It
therefore depends on physical properties of the sea ice layer and of the wave spectrum in that
particular portion of the sea.
Since the cost function $\Psi$ in Eq. (\ref{Psi}) is integrated over wave numbers, Eq. (\ref{hnu}) can depend on wave properties only in integrated form. Moreover, for small amplitude waves, nonlinear dependence of the stress on the wave strain, which would lead to an implicit dependence of $\nu$ on wave properties, can be ruled out.
We therefore disregard the dependence of $\nu$ on wave properties and
write
\beq
\beta=\beta(h)\ \Rightarrow\ \nu=\nu(h).
\label{beta(h)}
\eeq

As Eq. (\ref{hnu}) does not allow one to determine the
specific couple $(h,\nu)$ of ice thickness
and viscosity relevant to the selected sea region, the functional dependence
$\beta=\beta(h)$ remains undetermined.

An alternative representation of the GPI 
thickness distribution for a given SAR acquisition is thus required
to lift the undeterminacy in $(h,\nu)$.
A possibility is offered by the salt-flux model described in 
\cite{pedersen2004sea}, developed specifically to 
describe the formation, transport and desalinization of GPI in the Odden region of the Greenland 
Sea, and carefully validated with in situ measurements. 
Indeed, an oceanographic campaign was carried out into the Odden from March 3 to March 13 1997 on the R/V Jan Mayen\cite{pedersen2004sea}. The Odden locations visited during the cruise operations were imaged by the SAR onboard the ERS2 satellite. In particular, a couple of ERS2 SAR images gathered on March 9, 1997, in coincidence with the acquisition of a Datawell directional wave buoy at (73N, 1W)\cite{wadhams1997wave}, and on March 11, 1997, centered at $73.24^o$ N, $9.11^o$ W, were considered, respectively.

The equivalent solid ice thicknesses, predicted by the salt-flux model for the areas imaged by the 
concurrent ERS2 SAR images, were computed according to \eqref{heq}. 
Comparison of the salt-flux predictions by the model with
$in$ $situ$ ice samplings collected in coincidence with SAR data takes, yields, 
in the two dates selected, values of the equivalent ice thickness $h$ in the range 4-9 cm, 
with relative error $\Delta h / h\simeq0.26$.

\begin{figure}[htbp]
\centering
\includegraphics[width=\linewidth]{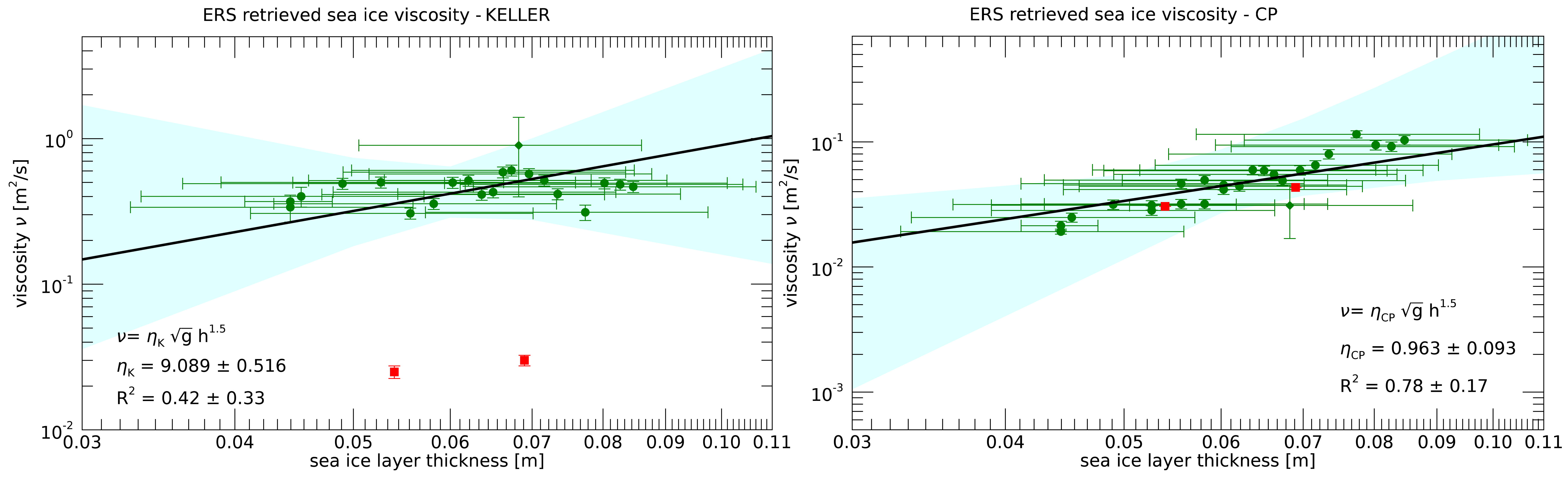}
\caption{GPI viscosity vs equivalent thickness obtained by running Keller's model (left panel) and CP model (right panel), for the calibration instances selected from the ERS2 SAR images gathered in the Odden region on 9 March and 11 March 1997, respectively. Viscosity values are given by the absolute minimum of the SAR cost function corresponding to the equivalent ice thicknesses predicted by the ice salt-flux ice model (Pedersen \& Coon, 2004). The two red points represent the pure grease ice viscosity grown in laboratory (Newyear \& Martin, 1999). Cyan areas represent the regions of 95\% confidence level for the best fit line obtained as Bayesian linear regression (Kelly, 2007).}
\label{fig:cal}
\end{figure}

A total of 24 windows at increasing distance from the ice edge were selected
for application of the SAR inversion scheme,
to sample the overall GPI thickness spatial variability predicted by 
the salt-flux model.
Figure \ref{fig:cal} shows the calibration data obtained for both Keller's and CP models.

GPI viscosity values were estimated after minimization of the SAR cost function, by imposing the GPI equivalent thicknesses predicted by the salt-flux model.
A Bayesian regression scheme \cite{kelly2007some} 
was adopted to account for the uncertainties in the ice viscosity and thickness, in the assumption of
a linear relation between ice viscosity and ice thickness on a log-log scale.
Figure \ref{fig:cal} shows the data used for the fitting procedure. A coefficient of determination equal to $R^2=0.42$ for the Keller's model and $R^2=0.78$ for the CP model, respectively, was found. The cyan areas in figure \ref{fig:cal} represent the region where the straight line is bounded within the 95\% confidence level of the linear fit. 

%

The data in Fig. \ref{fig:cal} do not allow us to pin down a precise form of constitutive relation
$\nu=\nu(h)$. We have chosen to give special weight to the laboratory data in \cite{newyear99}, red points in Fig. \ref{fig:cal}, which suggest us the empirical law
\begin{equation}
    \nu=\eta g^{1/2} h^{3/2}.
\label{cal_fun}
\end{equation}
The dimensionless constant $\eta$ may still depend on 
internal properties of the GPI such as the frazil and pancake size, and the water viscosity. Constant $\eta$ would describe a situation in which 
the ice response to the waves only depends on characteristic length 
$h$ and  characteristic time  $(h/g)^{1/2}$, with
Eq. (\ref{cal_fun}) providing the only dimensionally consistent expression for the viscosity dependent on the two parameters.
We point out that Eq. (\ref{cal_fun}) is conceptually different from Eq. (\ref{hnu}).
Indeed Eq. (\ref{cal_fun}) represents a constitutive relation for the GPI, while Eq. (\ref{hnu}) 
comes from the viscous model structure 
and does not carry any information on the ice rheology.

Substituting Eq. (\ref{cal_fun}) into Eqs. (\ref{hnu}) and 
(\ref{alphas}), we find
in the two cases of the Keller's and the CP model:
\begin{equation}
\beta_K=\eta_K g^{1/2} h^{5/2},
\qquad\qquad  
\beta_{CP}=\eta_{CP} g^{1/2} h^{-3/2}.
     \label{nudih}
\end{equation}



We can now carry out the calibration procedure 
by fitting the Odden data showed in Figure \ref{fig:cal}. The following values for the dimensionless coefficients $\eta_K$ and $\eta_{CP}$ are obtained:
\begin{equation}
    \eta_K=9.089\pm0.516,
\qquad\qquad 
    \eta_{CP}=0.963\pm0.093 
\label{eta_hat}. 
\end{equation}

The smallness of the uncertainty in $\eta_K$ and $\eta_{CP}$ is an indication that the 
constitutive relation Eq. (\ref{cal_fun}) provides a physically reasonable description of
the rheology of the ice layer.
The difference between $\eta_K$ and $\eta_{CP}$, however, is striking. 
The most natural explanation for this phenomenon is that the viscosity in the Keller model refers
to the whole GPI mixture, while in the CP model it refers properly to grease ice layer. 
In fact, the CP model provides GPI viscosities which are consistent with laboratory measurements
\cite{newyear1999comparison}, while in the case of the Keller's model the GPI viscosity is about one order of magnitude higher than the laboratory measurements.
We think that such higher viscosity, required to fit the data with the Keller's model, reflects the contribution of pancakes to the overall ice viscosity.
Once the value of $\beta_{\rm x}$ from the cost function of the actual SAR inverted tile is available, the average ice thickness of the GPI region travelled by the wave system can be computed from
Eq. (\ref{eta_hat}) into Eq. (\ref{nudih}). The result is
 \begin{equation}
 h=(0.414\pm0.009)g^{-1/5}\beta_K^{2/5} \label{acca_k}
 \end{equation}
 for the Keller's model, and
 \begin{equation} 
     h=(0.975\pm0.064)g^{1/3}\beta_{CP}^{-2/3} \label{acca}
 \end{equation}
for the CP model, respectively. The uncertainty of $\beta$ 
in Eqs. (\ref{acca_k}) and (\ref{acca}) is assumed negligible compared to that of 
 $\eta$.

\subsection*{Examples of SAR inferred GPI thicknesses}
Eqs. (\ref{acca_k}) and (\ref{acca}) are imposed in the SAR-wave inversion procedure that is applied to map the GPI thicknesses from a Sentinel-1A (S1A) SAR image gathered in the Beaufort Sea, and from three COSMO-SkyMed (CSK) images collected in the Weddell Sea, Antarctica, respectively. All details of SAR images are listed in Table \ref{tabella}.

\begin{table}[htbp]\scriptsize
\begin{tabular}{|c|c|c|c|c|c|c|c|c|c|c|c|c|c|}
\hline
\multicolumn{5}{|c|}{Satellite}                & \multicolumn{3}{c|}{\begin{tabular}[c]{@{}c@{}}Observed SAR spectral \\ parameters @ peak location\end{tabular}}        & \multicolumn{3}{c|}{\begin{tabular}[c]{@{}c@{}}Retrieved SAR spectral \\ parameters @ peak location\end{tabular}}        & \multicolumn{3}{c|}{\begin{tabular}[c]{@{}c@{}}Retrieved ocean wave \\ spectrum parameters\end{tabular}}               \\ 
\hline
\hline
\begin{tabular}[c]{@{}c@{}}Date/\\ Time\end{tabular}& 
\multicolumn{1}{l|}{Sensor} & 
\multicolumn{1}{l|}{Band/Pol} & 
\begin{tabular}[c]{@{}c@{}}Orbit/\\ look\end{tabular} & 
\begin{tabular}[c]{@{}c@{}}Pixel size\\  {[}m{]}\end{tabular} & \begin{tabular}[c]{@{}c@{}}Value\\  {[}arb. unit{]}\end{tabular} & \begin{tabular}[c]{@{}c@{}}Wavelength \\ {[}m{]}\end{tabular} & \begin{tabular}[c]{@{}c@{}}Dir. \\ {[}deg{]}\end{tabular} & \begin{tabular}[c]{@{}c@{}}Value\\  {[}arb. unit{]}\end{tabular} & \begin{tabular}[c]{@{}c@{}}Wavelength \\ {[}m{]}\end{tabular} & \begin{tabular}[c]{@{}c@{}}Dir. \\ {[}deg{]}\end{tabular} & \begin{tabular}[c]{@{}c@{}}Wavelength \\ {[}m{]}\end{tabular} & \begin{tabular}[c]{@{}c@{}}Dir.\\ {[}deg{]}\end{tabular} & Hs {[}m{]} \\ \hline
\begin{tabular}[c]{@{}c@{}}11/1/2015 \\ 17:23\end{tabular}
& S1A   & C/HH  &
\begin{tabular}[c]{@{}c@{}}DESC/\\ RIGHT\end{tabular}        & 10  
& 13   & 113   & 225 & 39 & 118  & 214  & 118    & 214  & 1.45       \\ \hline
\begin{tabular}[c]{@{}c@{}}3/21/2019 \\ 17:16\end{tabular}
& CSK1 & X/VV & 
\begin{tabular}[c]{@{}c@{}}DESC/\\ RIGHT\end{tabular} & 15
& 27  & 100  & 133 & 25 & 108 & 106  & 128 & 274 & 2.00       \\ 
\hline
\begin{tabular}[c]{@{}c@{}}3/22/2019 \\ 17:10\end{tabular}
& CSK4 & X/VV & 
\begin{tabular}[c]{@{}c@{}}DESC/\\ RIGHT\end{tabular} & 15                     
& 24  & 110  & 114 & 25 & 121 & 108  & 147 & 274 & 2.11       \\ 
\hline
\begin{tabular}[c]{@{}c@{}}3/30/2019 \\ 10:00\end{tabular}
& CSK4 & X/VV & 
\begin{tabular}[c]{@{}c@{}}ASC/\\ RIGHT\end{tabular} & 15                  
& 32  & 141  & 234 & 26 & 141 & 234  & 168 & 232 & 1.90       \\ 
\hline
\end{tabular}
\caption{Characteristics of the Sentinel-1A and COSMO-SkyMed SAR images selected for the application the SAR inversion procedure to infer GPI thickness. Results of the best fit SAR image spectra at the peak location along with the retrieved ocean wave spectra parameters in open sea are also reported.}
\label{tabella}
\end{table}

\begin{figure}[htbp]
\centering
\includegraphics[width=0.95\linewidth]{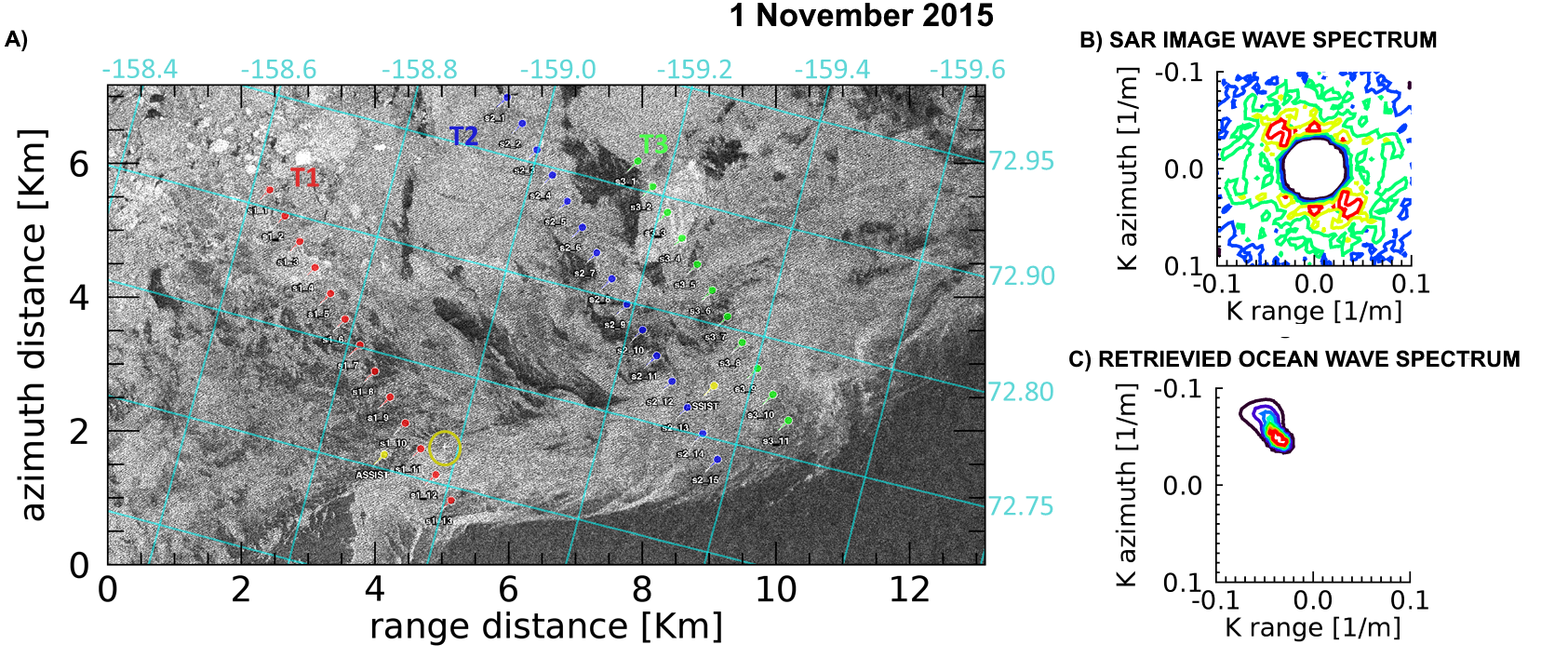}
\caption{A) Portion of the Sentinel-1A (S1A) IW-mode SAR image acquired on 1 November 2015 at 17:23 UTC in the Beaufort Sea. Representation is in the ground range/azimuth SAR acquisition reference frame. Superimposed is the lat/lon grid. T1 (red), T2 (blue), and T3 (green) are the three transects selected for ice thickness estimation around the corresponding points. The yellow pins indicate the location of ice measurements made according to the ASSIST protocol within $\pm$1 hour the SAR image data take. The bright cross encircled by the yellow ring is the imaged R/V Sikuliaq. B) Observed SAR image wavenumber spectrum in open sea at 72.68$^o$ N 159.20$^o$ W. C) Corresponding retrieved ocean wavenumber directional spectrum.}
\label{fig:1nov}
\end{figure}

\subsubsection*{Arctic – Beaufort Sea}
On 1 November 2015, an IW-mode S1A SAR image was collected in the advancing MIZ of the Beaufort Sea (Figure \ref{fig:1nov}). At the time of SAR acquisition, the R/V Sikuliaq was in operation to carry out a field cruise as part of a large collaborative program to study the coupled air-ice-ocean-wave processes occurring in the Arctic during the autumn ice advance\cite{thomson2018overview}.

A sharp ice edge can be detected in the S1A SAR image separating the open sea (bottom part) from the ice field. The region of icefield directly exposed to open sea appears to be mainly composed of GPI. The hourly visual observations conducted aboard the R/V Sikuliaq using the ASSIST protocol  (http://www.iarc.uaf.edu/icewatch) recognized three ice types, namely, young grey ice, pancakes and first year ice, along with an estimation of the primary thickness and partial concentrations for each ice type. Occasional ship-side retrieval of frazil ice and pancakes samples was also carried out as a concurrent measurement, and a number of directional wave buoys were deployed. 

At the time of SAR acquisition, a SWIFT wave buoy was floating in open sea around the ice edge, close to the ship, where an incoming wave spectrum was measured with wave height $H_S\simeq1.2$ m. Figure \ref{fig:1nov}, panel B, shows the two-dimensional SAR image wavenumber spectrum observed by S1A in a $512\times512$ pixels open sea area, centered at (72.68N, 159.20W), close to the ice edge and near to the SWIFT buoy location; the corresponding two-dimensional directional ocean wave spectrum, shown in the panel C, is obtained using the SWIFT buoy wave spectrum as first guess input\cite{hasselmann1991nonlinear}. The SAR inversion procedure returns a consistent wave directional spectrum, with a slightly higher value of wave height $H_S\simeq1.5$ m (Table \ref{tabella}).

\begin{figure}[htbp]
\centering
\includegraphics[width=0.7\linewidth]{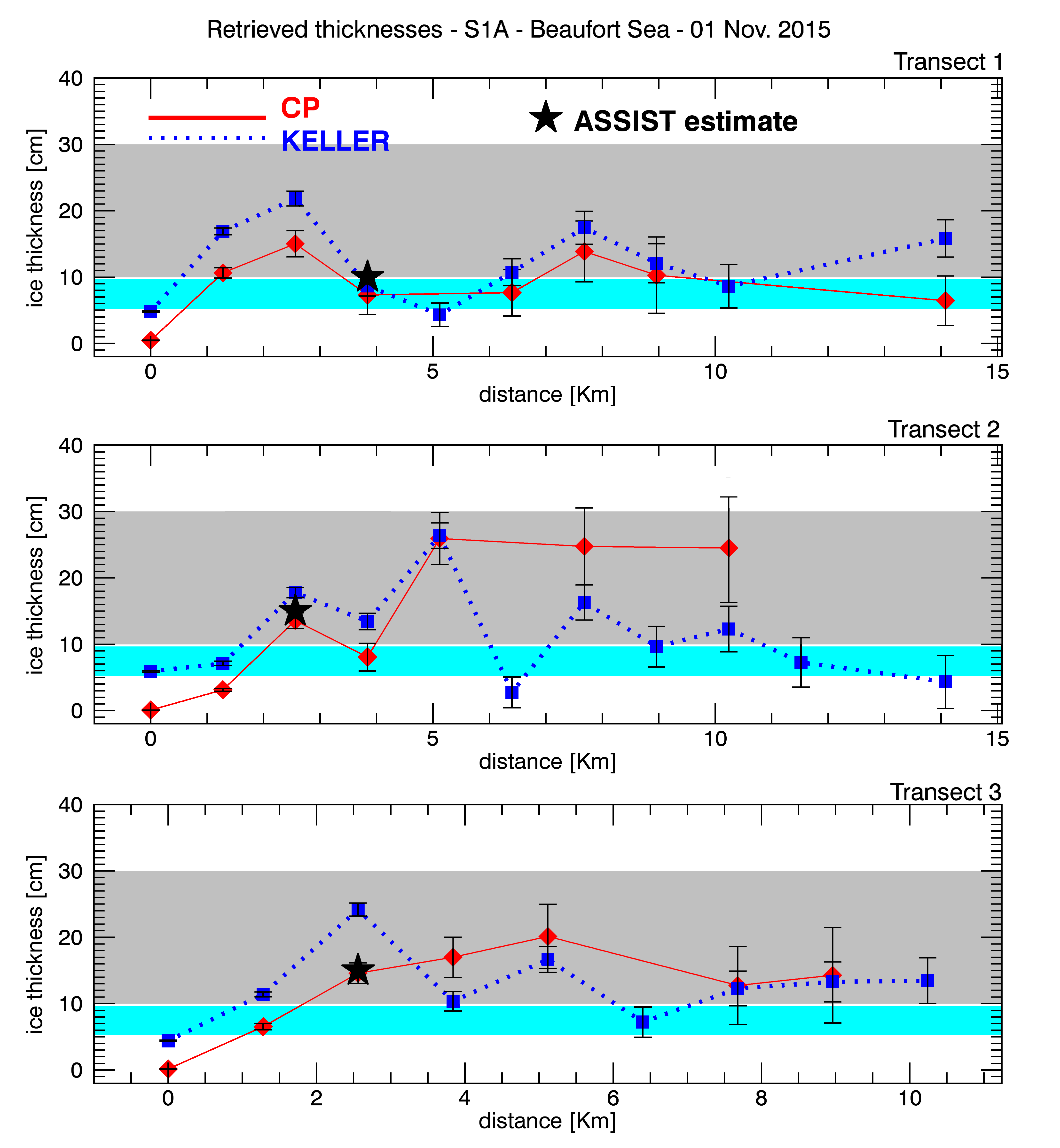}
\caption{Estimated sea ice thicknesses over the three transects selected on the S1A SAR image using the Keller's (blue dots) and CP model (red dots), respectively. The cyan band represents the variability of the SMOS thicknesses in the area covered by the transects. The grey band represents the range of thicknesses of the sea ice cover using the ASSIST protocol on 1 November. The "star" symbol represents the primary thickness estimate using the ASSIST protocol at the location of the corresponding SAR window within one hour the SAR acquisition.}
 \label{fig:invi1nov}
 \end{figure}

Following the dominant wave direction, three transects were selected, respectively formed by 13 (transect 1, T1), 15 (T2) and 11 (T3) SAR image tiles of size $256\times256$ pixels, as shown in Figure \ref{fig:1nov}. The retrieved ice thicknesses are shown in Figure \ref{fig:invi1nov}, along with the range of thicknesses from both the ASSIST protocol observations.

Results were also compared with the sea ice  thickness retrieved from  the L-band microwave sensor SMOS (Soil Moisture and Ocean Salinity).  Daily coverage of the polar regions with a resolution of about 35 km $\times$ 35 km are indeed inferred through an empirical method \cite{huntemann2014empirical} from SMOS acquisitions. It is worth to note that SMOS sea ice thicknesses (cyan band in Figure \ref{fig:invi1nov}) underestimates the primary sea ice thicknesses from ASSIST protocol (grey band in Figure \ref{fig:invi1nov}). 

Thicknesses from SAR retrievals using CP and Keller's models are all within the SMOS and ASSIST range of thicknesses and show a consistent behaviour running over the ice field depth. The overall trend shows increasing thicknesses going from the ice edge up to values close to 25 cm.

In transect T1 a good agreement between CP and Keller results is observed. At the fourth location, close to the ship, a thickness of $7.3\pm2.9$ cm from CP model and $8.7\pm1.5$ cm from Keller's model are estimated, which are consistent with the 10 cm of primary thickness from ASSIST estimation as well as with the average thickness $9.6\pm3.8$ cm obtained from pancake measurements collected in six recovery locations \cite{wadhams2018pancake}.

In transect T2, thicknesses retrieved with both CP and Keller's models agree for the first 5 km inside the ice fields (windows 1--5). A good correspondence between ASSIST thickness (15 cm) and SAR retrievals (CP: $13.6\pm1.2$ cm; Keller: $17.8\pm0.8$ cm), at the third location, is again observed. From window 6 onward, the thicknesses inferred with the two models show different trends. This could be explained by inspecting the SAR image in Figure \ref{fig:1nov}. In correspondence to window 6, a dark zone is indeed observed, which is consistent with grease ice feature, possibly mixed with water, in the absence of pancakes. In such area wind can easily transfer energy to waves, altering the attenuation trend by the GPI, measured up to window 5.

In this regard, we point out that the SAR inversion procedure compares the open water wave spectrum with the wave spectrum observed at the specific window in sea ice. What the SAR procedure actually returns is therefore the average $h^*_n$ of the effective thickness $h$ from the open sea to the given window,
\begin{equation}
    h^*_n=\frac{1}{n}\sum_{i=1}^nh_i.
    \label{hmed0}
\end{equation}
In order to find the GPI thickness specific only for the window $n$, the contribution from the previous windows has then to be removed as follows,
\begin{equation}
h_n=nh^*_n-(n-1)h^*_{n-1}.
\label{hmed}
\end{equation}
Negative values for $h_n$ can result. In this case, the data are not reported, as it happens for CP thickness estimation of window 6 belonging to T2. 
Moreover, note that the CP model specifically represents a systems in which objects, identified as pancakes or small ice floes, float upon a viscous layer, i.e. grease ice. 
A negative $h_i$ may therefore indicate an absence of pancakes in the considered location, which could lead to wave build-up from wind action.

The two models return thickness values in the following windows of T2 that vary somewhat from window to window, even though they remain in the range defined by the SMOS and ASSIST estimates. This difference in thicknesses could reveal changes of the ice cover, which cannot longer be handled by the CP model.

Finally, as for T1, in transect T3, CP and Keller's model infer comparable values of the
thickness. Also in the third location, there is good agreement between ASSIST thickness (15 cm) and CP thickness ($14.6\pm1.5$ cm) from SAR, but a slight disagreement with Keller's thickness, which resulted as high as $24.2\pm1.0$ cm.

\subsubsection*{Antarctica – Weddell Sea}
In the context of the Year of Polar Prediction (YOPP) initiative of the World Meteorological Program, \cite{jung2016advancing} a survey of the advancing ice edge of the Weddell Sea was performed during March 2019 by exploiting the imaging capability of the four SAR satellites forming the COSMO-SkyMed (CSK) constellation. We have applied the SAR-wave inversion technique to the SAR WIDEREGION images acquired on 21, 22 and 30 March 2019, where at least one CSK SAR satellite was able to image the sea ice edge.

A steady open ocean wave field of $H_S\simeq2.0$ m carrying dominant wavelengths ranging from $\lambda\simeq128$ m to $\lambda\simeq168$ m is observed to cross the GPI fields in the three dates. The 
associated wave spectra in open ocean are obtained from SAR inversion using the closest 2D WAM spectrum \cite{p16951} available for each date. Details of SAR images characteristics and retrieved wave fields are listed in Table \ref{tabella}.
 
  \begin{figure}[htbp]
 \centering
 \includegraphics[width=0.95\linewidth]{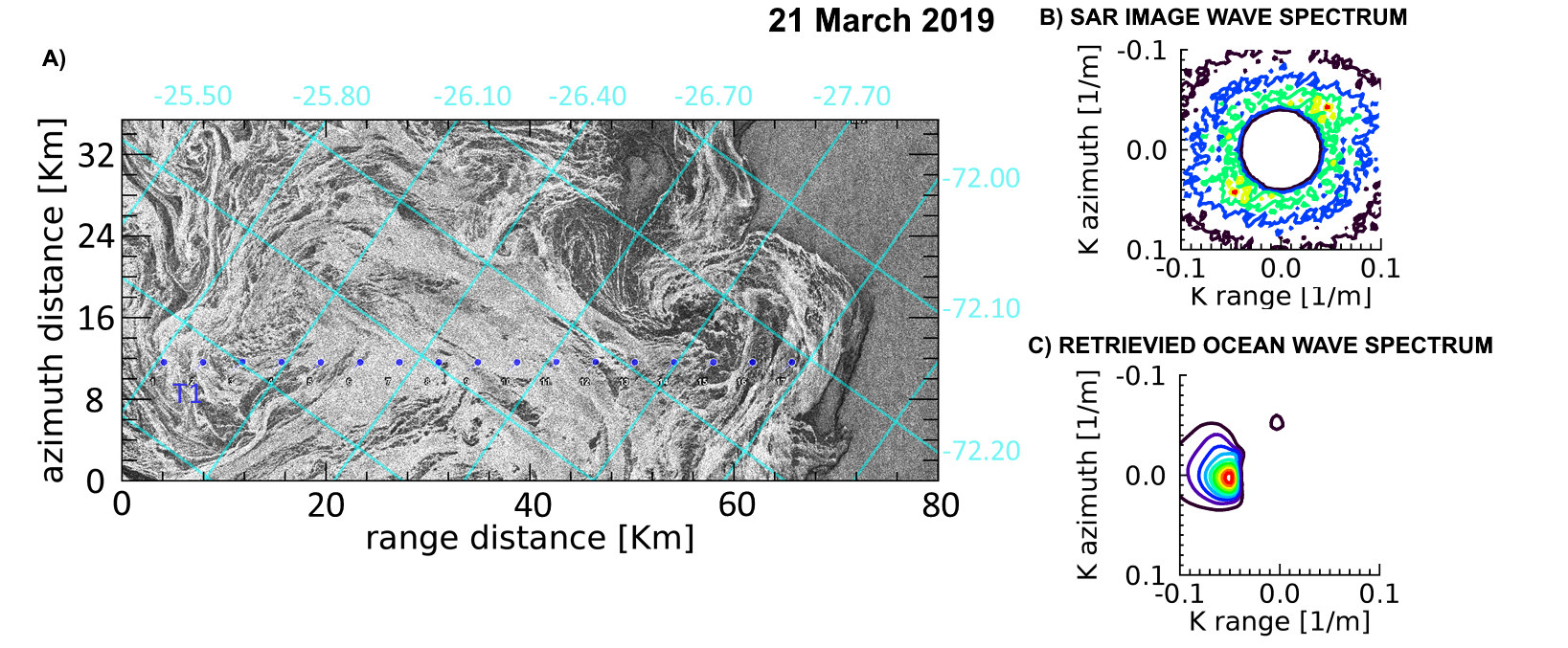}
 \caption{A) Portion of the COSMO-SkyMed Wide Region mode SAR image acquired in the Weddell Sea on 21 March 2019, represented in the geometry of the SAR acquisition reference frame. Superimposed are the lat/lon grids. The aligned blue dots represent the locations along the transect T1 selected for ice thickness estimation. B) Observed SAR image wavenumber spectrum in open sea at (72.00 S, 27.15 W). C) Corresponding retrieved wavenumber directional spectrum in open ocean.}
 \label{fig:21mar}
 \end{figure}

\begin{figure}[htbp]
\centering
\includegraphics[width=0.95\linewidth]{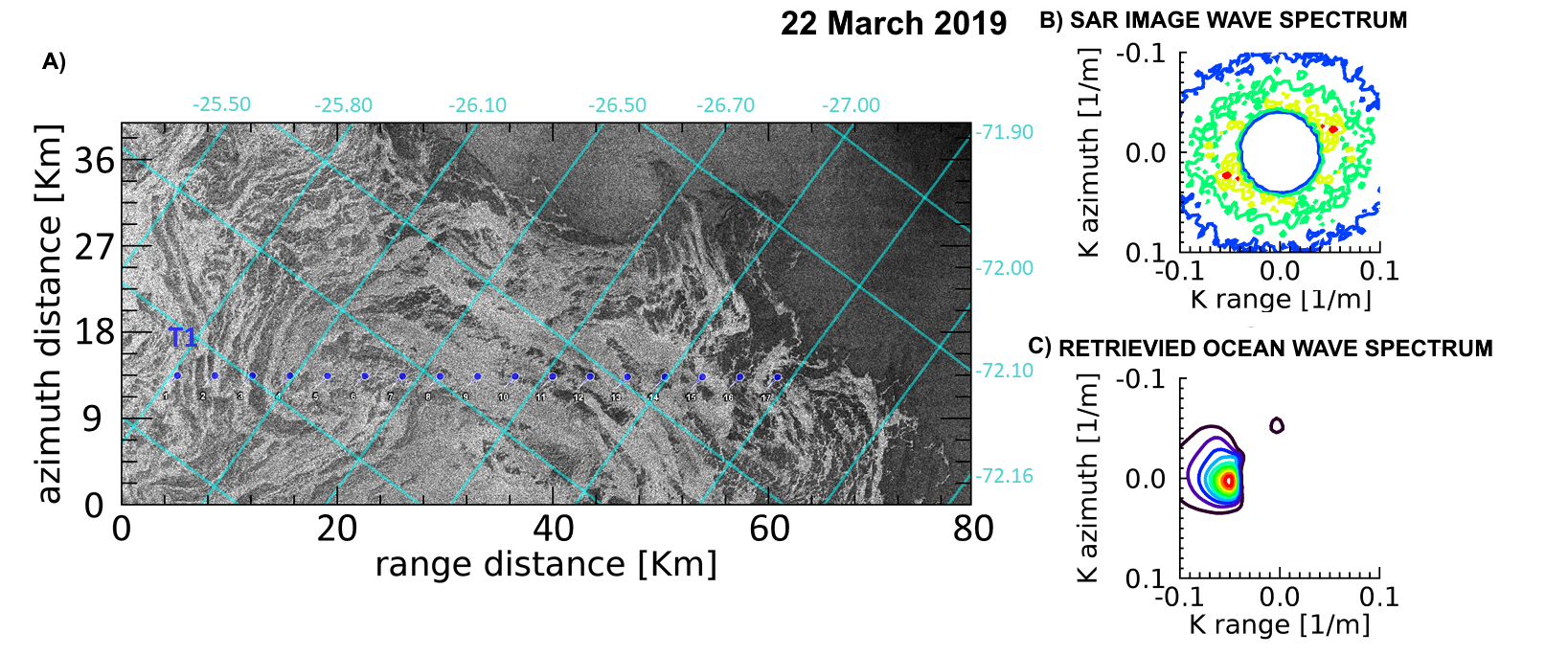}
\caption{A) Portion of the COSMO-SkyMed Wide Region mode SAR image acquired in the Weddell Sea on 22 March 2019, represented in the geometry of the SAR acquisition reference frame. Superimposed are the lat/lon grids. The aligned blue dots represent the locations along the transect T1 selected for ice thickness estimation. B) Observed SAR image wavenumber spectrum in open sea at (72.00 S, 25.65 W). C) Corresponding retrieved wavenumber directional spectrum in open ocean.}
\label{fig:22mar}
\end{figure}
 
\begin{figure}[htbp]
\centering
\includegraphics[width=0.95\linewidth]{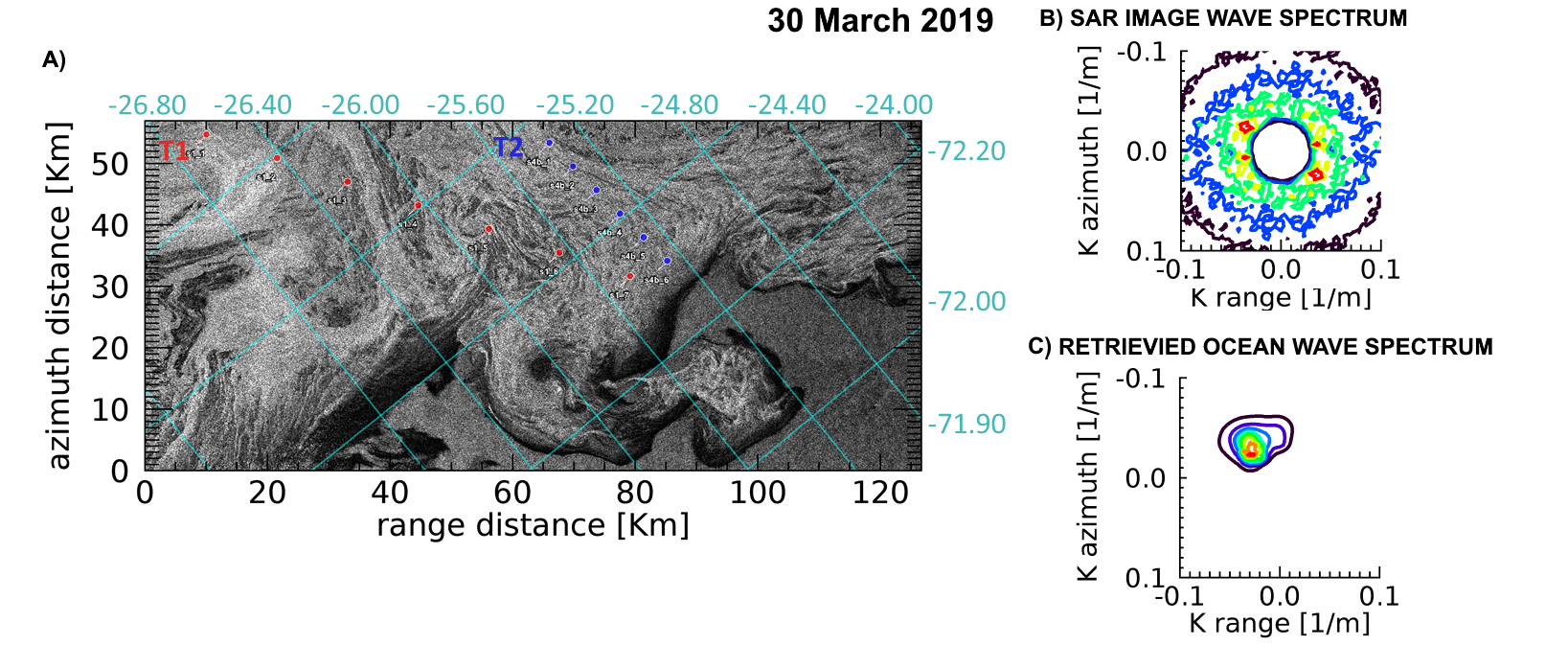}
\caption{A) Portion of the COSMO-SkyMed Wide Region mode SAR image acquired in the Weddell Sea on 30 March 2019, represented in the geometry of the SAR acquisition reference frame. Superimposed are the lat/lon grids. The aligned red (blue) dots represent the locations along the transect T1 (T2) selected for ice thickness estimation. B) Observed SAR image wavenumber spectra in open sea at (71.95 S, 25.92 W). C) Corresponding retrieved wavenumber directional spectrum in open ocean.}
\label{fig:30mar}
\end{figure}

\begin{figure}[ht]
\centering
 \includegraphics[width=0.95\linewidth]{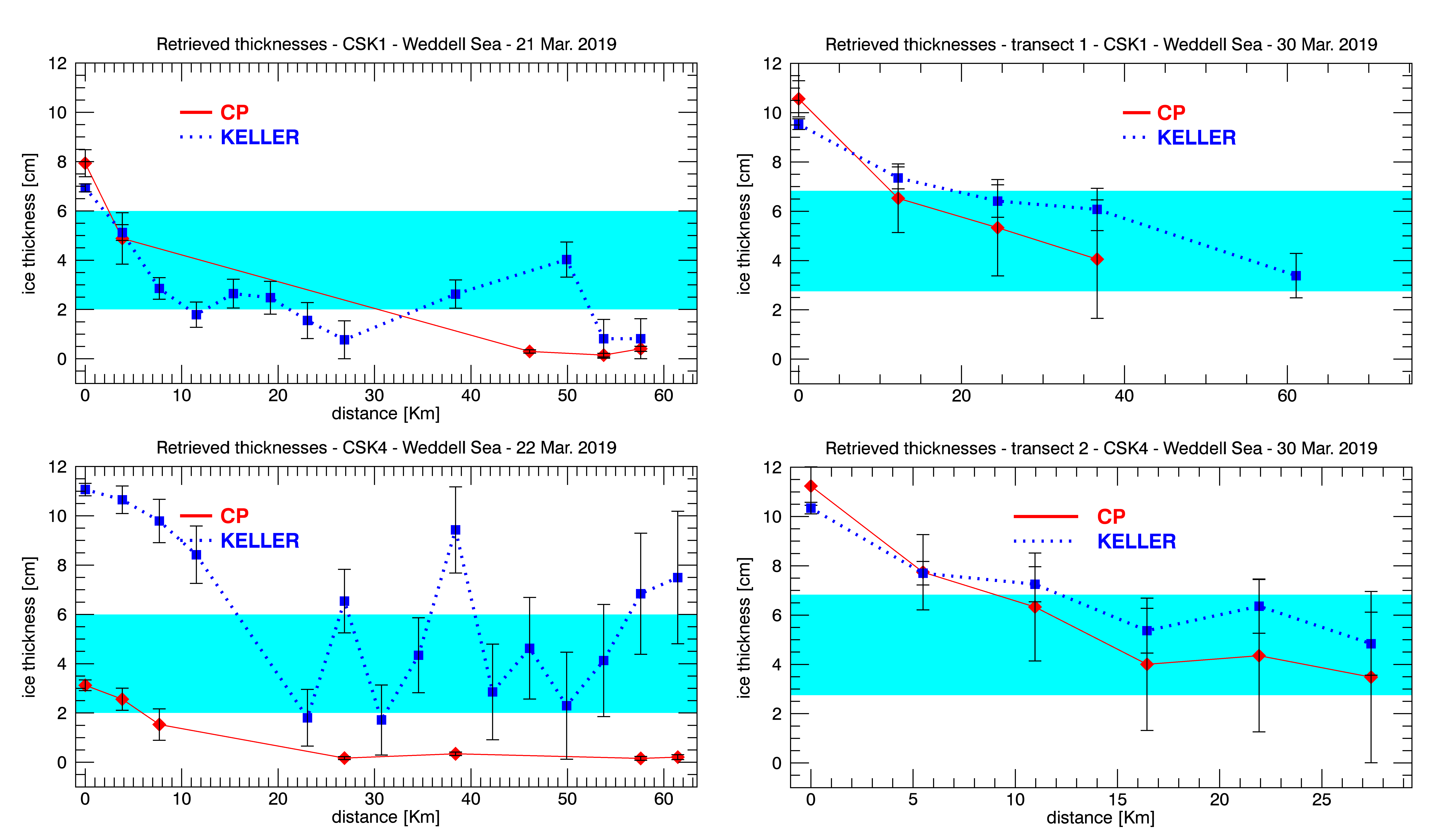}
\caption{As for Figure \ref{fig:invi1nov} but relevant to the transects selected on the CSK SAR images acquired in March 2019.}
\label{fig:invimar}
\end{figure}

A common feature of the three SAR images is represented by the dark areas adjacent to the ice edge at direct contact with the open sea, which reveal the presence of bands of pure frazil/grease ice, from which pancakes originate. The SAR images taken on 21 and 22 March almost overlapped the same area of the Weddell Sea and were able to image the outermost part of the developing GPI field; on 30 March, a wider area was imaged by the SAR instrument, which includes the region imaged on 21 ad 22 March, thus showing an expansion of the GPI field.
Figures \ref{fig:21mar}, \ref{fig:22mar} and \ref{fig:30mar} show the interfaces separating open sea with the developed GPI fields imaged by the CSK SAR images where the SAR-wave inversion procedure was applied.
As for Figure \ref{fig:1nov}, transects are selected approximately following the propagation of the dominant wave. For each location of the transect the estimated thicknesses are plotted in Figure \ref{fig:invimar} according to the distance inside the ice field. The SMOS thicknesses over the area covered by the transects are also reported as band of variability.

As a general comment, 
smaller values of the thickness, much closer to the SMOS estimations,
resulted from the SAR inversion,
compared to the Beaufort Sea case study.
In contrast to the case study in Arctic, the region closer to the ice edge, showed thicker ice up to about 3 cm higher the extremal SMOS value, with a trend to decrease when running towards the inside of the icefield. The phenomenon could be caused the action of the wave field that tends to compact the GPI in the region immediately surrounding the ice edge.

Figures \ref{fig:21mar} and \ref{fig:22mar} show a very dynamic ice field with many darker areas of pure grease ice, and possibly water, embedded in the GPI environment. As occurred in
transect T2 in Figure \ref{fig:invi1nov}, also in this case the CP model often appears unable 
to achieve
thickness retrieval. It is worthwhile to note that for the 21$^{\rm st}$ of March a general trend of decreasing thickness is obtained using both models, with comparable values of $h$, when available. Similarly, a general agreement between the inferred thicknesses is observed for the two transects of the 30$^{\rm th}$ of March. 

\section*{Discussion}
A new technique aimed at the estimation of GPI thickness from the inversion of SAR image wave 
spectra is proposed. The intrinsic undetermination of any SAR inversion technique based on a 
viscous layer model is resolved with an empirically determined constitutive relationship 
between the viscosity and the thickness of the GPI layer, $\nu\propto h^{3/2}$.

A similar power-law relation, $\nu\propto h^2$, was proposed in 
\cite{sutherland2019two} on dimensional grounds, based on the hypothesis that
$\nu$ only depends on the wave frequency and the ice thickness. We point out that in 
our case $h$ is an effective thickness, proportional through Eq. (\ref{heq})
to the ice volume fraction $C$. The ice volume fraction $C$ is expected to increase with $h$
under the effect of the buoyancy pressure\cite{olla18}, which is itself proportional to the
effective ice thickness $h$, $P_s\sim\Delta\rho hg$, where $\Delta\rho$ is the mass density
gap between ice and liquid water. This gives a physical content to Eq. (\ref{cal_fun}), which
goes well beyond the level of a dimensional relation.  The microscopic justification 
of Eq. (\ref{cal_fun}), however, remains elusive, buried in the dependence
of the dimensionless parameter $\eta$ in Eq. (\ref{cal_fun}) on internal GPI properties,
such as the geometry and size of frazil crystals and pancakes, and the water viscosity $\nu_w$.
We point out that any dependence of $\eta$ on $\nu_w$, for dimensional consistency, would
require dependence of $\eta$ on an additional time scales, which would require in turn dependence
of $\eta$ on $g$ or other mechanical parameters, possibly associated with the stress in 
the ice matrix. In all cases, a picture of the GPI more akin to a brittle solid or a granular
medium, than to a fluid suspension\cite{decarolis05}, is suggested.

The prefactor in the constitutive equation is determined through a calibration procedure for two different viscous layer models that approximate wave propagation in GPI: the CP and the Keller's model. To achieve the task, external thickness information provided by the salt-flux model related to the 1997 Odden Ice Tongue \cite{pedersen2004sea} are used as reference data.

The GPI viscosity-thickness relationship obtained for the CP model is comparable with the one found for grease ice grown in wave tank at comparable thicknesses \cite{newyear1999comparison}. This is an encouraging result, as the CP model assumes that viscous effects on wave propagation are due to grease ice only.
On the other hand, the Keller's model envisions a layer of grease ice in which pancakes and small floes are suspended. Therefore, the resulting effective viscosity is determined by all types of ice composing the ice layer. A physical interpretation for Eq.\ref{eta_hat} is possible following Mooney\cite{mooney1951viscosity}. Indicating with $\phi$ the volume fraction of the pancakes (suspensions) and with $\phi_c=\pi/6 \approx 0.52$ its value in the case of maximally packed spheres on the sites of cubic lattice, the effective viscosity of the mixture, $\nu$, should obey the following equation:
\begin{equation}
    \ln{\cfrac{\nu}{\nu_{\rm grease}}}=\cfrac{2.5\phi}{1-\phi/\phi_c}
\end{equation}
where for $\nu_{\rm grease}$ is taken the viscosity calibrated for CP. The values obtained for $\eta_{\rm K}$ and $\eta_{\rm CP}$ in Eq. (\ref{eta_hat}) are compatible with a layer in which the volume fraction of the pancakes is about $\phi=0.33$. This value of concentration is physically sound and consistent with field measurements reported for the Weddell Sea in Antarctica\cite{doble2003pancake}.


The application of the two calibrated viscosities to different SAR images reveals that the inferred thicknesses obtained by the two models are generally in good agreement, and agree with the other estimates of GPI thickness available. The CP model seems to work better in real GPI fields, when both grease and pancake ice are present; the Keller's model seems more robust and to work well also in pancake free situations.

\section*{Methods}


\subsection*{The SAR-wave inversion procedure}\label{sec:SAR}
The approach considers the ocean waves generated in open sea that cross the ice edge and finally propagate in the icefield. It is assumed that the wave spectral features are modified by the GPI cover through viscous interactions modeled either by the CP or Keller's model. A SAR cost function is defined to measure the distance between the observed SAR image spectrum and the one simulated as a function of the wave models' parameterization, i.e., ice thickness and viscosity.

\subsubsection*{SAR inversion in open ocean}
The first step consists of the estimation of the directional open ocean wave spectrum at the boundary of the icefield from the available SAR image. The corresponding wavenumber SAR image spectrum is estimated from a tile, typically 514 x 512 pixels in range and azimuth. If the single look complex SAR image product is available, the SAR image cross-spectrum is calculated starting from the temporal separation between SAR looks, $\tau$. SAR cross spectra significantly reduce the speckle noise level while preserving the spectral shape, and provides information about the wave propagation direction \cite{engen1995sar}. The simulated SAR image spectrum is computed using the closed form expression of the non-linear ocean-to-SAR spectral form described in Hasselmann \& Hasselmann (1991) \cite{hasselmann1991nonlinear} and later modified for the SAR image cross-spectral transform \cite{engen1995sar}. For $\tau=0$ the SAR image cross-spectrum becomes a real-valued quantity and reduces to the standard multilooked SAR image spectrum. According to the procedure proposed by Mastenbroek \& De Valk (2000)\cite{mastenbroek2000semiparametric}, the wind-generated ocean spectrum is first estimated. The parametric representation given by Donelan et al. (1985)\cite{donelan1985directional} is assumed, which depends on the inverse age of the dominant wave, $\Omega = U_{10}/c_p$ , where $U_{10}$ is the 10 m asl wind speed and $C_p$ is the phase velocity of the dominant wave, and the wind vector as well. The latter can be taken either from in situ measurements, if available, or SAR-estimated using the wind vector from a numerical weather prediction model, i.e. ECMWF, as first guess \cite{portabella2002toward}. Then, the residual wave spectrum is estimated by assuming a parametric representation of directional swells according to the JONSWAP-Glenn spectral shape coupled with a directional spreading function of the Mitsuyasu type, properly extended for swell propagation \cite{goda2010random}. This choice accounts for non-linearity in SAR imaging features induced by high swells that can occur in the polar seas. Thus, the implemented SAR retrieval of swells adopted a non-linear retrieval scheme which minimizes a cost function with the truncated Newton method implemented in IDL© with respect to the following seven parameters: the dominant wind wavenumber vector; the dominant swell wavenumber vector; the swell wave height; the shape parameters and the directional spread of the swell distribution. 
The Hasselmann \& Hasselmann (1991)\cite{hasselmann1991nonlinear} inversion method 
is applied in alternative, whenever the directional wave buoy data is available.

\subsubsection*{SAR inversion in the GPI field}
A series of adjacent SAR tiles of 256x256 pixels centered at the position $\bf{x}$ in the SAR reference frame and running from the ice edge along a straight line parallel to the direction of the incoming dominant wave are selected through the GPI field. For each of them, the SAR image (cross-)spectrum is computed. For the SAR images analyzed in this paper, the pixel sizes are 12.5 m for ERS2, 10 m for Sentinel-1A and 15 m for COSMO-SkyMed SAR images. Thus, the spatial resolution of the final GPI thickness and viscosity estimates is of the order of $\approx 10 {\rm Km}^2$. It is assumed that the GPI cover alters the incoming ocean wave spectrum $S_w({\bf k})$ according to the following expression \cite{wadhams2004sar,sutherland2016airborne}:
\begin{equation}
S_i({\bf x},{\bf k};h,\nu,\gamma)=\tilde S_w({\bf k})\exp[-2q(h,\nu,\gamma)\Delta({\bf e}_{\bf k},{\bf x})]
\label{S_i}
\end{equation}
where $\Delta({\bf e}_{\bf k},{\bf x})$ is the distance traveled from the ice edge to the position $\bf{x}$ by the wave of wavenumber $\bf{k}$ heading the direction ${\bf e}_{\bf k}$. 
For each window located at position $\bf{x}$ into the SAR image, the SAR inversion scheme computes the waves-in-ice spectrum that minimizes the cost function defined as:
\begin{equation}
\Psi(\tau,{\bf x};h,\nu,\gamma)=\int [\Re(\tilde P_i(\tau;{\bf x},{\bf k})-P_i(\tau;h,\nu,\gamma;{\bf x},{\bf k}))]^2{\rm d}^2k,
\label{Psi}
\end{equation}
where $\tilde P_i$ and $P_i$ are the observed and modelled SAR image spectra at $\bf{x}$ in the ice region and $\Re$ stands for the real part of the argument.
After the ocean wave spectrum has been modified according to Eq. (\ref{S_i}) through the selected wave model, the simulated SAR image spectrum is computed according to Hasselmann \& Hasselmann \cite{hasselmann1991nonlinear} and Engen \& Johnsen \cite{engen1995sar} for $\tau\neq0$.

\subsection*{The Close Packing (CP) model}\label{sec:CP}
The CP model envision the ice-covered ocean as a three-layer system, with the pancakes confined in the top layer, grease ice in the middle and ice-free water in the bottom. The ice cover produces both wave attenuation and wavelength decrease. The process is controlled by three parameters: the effective viscosity $\nu$ and the thickness $h$ of the grease ice, which is 
modeled as a continuous homogeneous medium, and a parameter, hereafter called $\gamma$, accounting 
for the effect of the pancakes. 
Small $\gamma$ describes a situation with widely spaced pancakes, 
passively transported by the field of the waves. 
Pancake spacing decreases with increasing $\gamma$, 
with $\gamma \gtrsim 7$ giving a threshold above which pancakes are
maximally packed (whence the name of the model). For $\gamma\gtrsim 7$, the pancake layer
becomes essentially inextensible. Field measurements by \cite{smith2019pancake} support the modelling assumption. In general, the pancake layer acts as an additional source of tangential stress on the grease ice

The dispersion relation for the model is obtained by imposing continuity of the stresses at the interface between the layers. We can obtain a simpler form by introducing dimensionless parameters:
\begin{equation}
    \hat{\nu}=\cfrac{k_{\infty}^{3/2}}{g^{1/2}}\nu,\qquad \psi=\frac{k_{\infty}^{1/4}g^{1/4}}{\nu^{1/2}}h,\label{nu_psi}
\end{equation}
where $k_{\infty}=\omega^2/g$ is the wavenumber in open sea, $\omega$ is the wave frequency, $g=9.8 {\rm m/s^2}$ is the gravitational acceleration and $k$ is the wave vector in the ice covered region.
For waves  at the peak of the wave spectrum ($k_{\infty}\approx 0.05 {\rm m^{-1}}$),  and for laboratory values of the grease viscosity ($\nu\approx 0.05 {\rm m^2/s}$), $\hat{\nu}$ is small ($\hat{\nu}\approx 5\times10^{-5}$).  For small $\hat{\nu}$ and generic $\gamma$ the dispersion relation can be written in the simple form
\begin{equation}
    \frac{k}{k_{\infty}}\simeq\frac{{\rm i}\hat{\alpha}\hat{\rho}\gamma}{1+\gamma}\hat{\nu}^{1/2}\tanh{(\hat{\alpha}\psi)},\label{cp}
\end{equation}
where $\hat{\alpha}=\sqrt{-{\rm i}}$ and $\hat{\rho}$ is the ratio of ice to liquid water density.
In the limit $\gamma\to 0$, corresponding to a pancake-free situation, the Keller's model is recovered:
\begin{equation}
    \frac{k}{k_{\infty}}\simeq1+8\hat{\rho}\hat{\nu}^{3/2}\left[{\rm i}\psi+\hat{\alpha}\frac{\cosh{\hat{\alpha}}\psi-1}{\sinh{\hat{\alpha}\psi}}\right].\label{keller}
\end{equation}
Both Keller's and CP models could in principle be utilized to describe waves in GPI, the main difference being the fact that in the case of the Keller’s model, $\nu$ accounts for the effective viscosity not just of grease ice, but of the whole mixture of pancakes and frazil crystals. 
For finite $\gamma$, the effect of ice on wave propagation is $O(\hat{\nu}^{1/2})$, while for $\gamma=0$, is $O(\hat{\nu})$, which tells us that the effect of pancakes is not a correction to Keller's model.
For small enough $\psi$ (i.e. small enough $h$), we can Taylor expand Eqs.(\ref{cp}) and (\ref{keller}). Back to dimension units we find:
\begin{eqnarray}
{\rm CP:}&\qquad& k\approx k_{\infty}+\hat{\rho}hk_{\infty}^2+\frac{{\rm i}\hat{\rho}g^{1/2}k_{\infty}^{5/2}}{3}\frac{h^3}{\nu},\label{cp_s}\\
{\rm Keller:}&\qquad& k\approx k_{\infty}+4{\rm i}\frac{\hat{\rho}k_{\infty}^{7/2}}{g^{1/2}}h \nu\label{keller_s}. 
\end{eqnarray}
We can identify in Eq. (\ref{cp_s}) a dispersion contribution  that reproduces the result by the mass-loading model\cite{peters1950effect}. The wave attenuation is the imaginary part of the wavenumber vector, $q=\Im(k)$,
that has in the two cases the simple power law form $q\propto h^3/\nu$ (CP model) and $q\propto h\nu$ (Keller's model).

\subsubsection*{Cost function Profiles for small ice thickness}
For small $h$ the asymptotic relations Eqs. (\ref{cp_s}) and (\ref{keller_s}) can be rewritten as
\begin{equation}
q(h,\nu,\gamma;\omega)=A(h,\nu)B(\omega),
\label{q(}
\end{equation}
where for the Keller's  model ($\gamma=0$)
\begin{equation}
    A(h,\nu)=h \nu,\qquad B(\omega)= 4\frac{\hat{\rho}k_{\infty}^{7/2}(\omega)}{g^{1/2}},\label{ABkeller}
\end{equation}
and for the CP model
\begin{equation}
A(h,\nu,\gamma)=\frac{h^3}{\nu},\qquad B(\omega,\gamma)=\frac{\gamma\hat{\rho}g^{1/2}k_{\infty}^{5/2}(\omega)}{3(1+\gamma)}.\label{ABcp}
\end{equation}
Substituting Eq. (\ref{q(}) into Eq. (\ref{S_i}) yields 
\begin{equation}
S_i({\bf x},{\bf k}_\infty;h,\nu,\gamma)=\tilde S_w({\bf k}_\infty)\exp[-2A(h,\nu)B(\omega({\bf k}_\infty),\gamma)\Delta({\bf e}_{\bf k},{\bf x})],
\end{equation}
which tells us that the cost function $\Psi$ in Eq. (\ref{Psi}) depends on the variables $h$ and $\nu$ only through $A(h,\nu)$:
\begin{equation}
\Psi(\tau,{\bf x};h,\nu,\gamma)=\Psi(\tau,{\bf x};A(h,\nu),\gamma).    
\end{equation}
The minimum of the cost function is determined by imposing
\begin{equation}
0=\nabla_{\nu,h}\Psi=\frac{\partial\Psi}{\partial A}\nabla_{\nu,h}A,
\nonumber
\end{equation}
whose solution can be written in the form
\begin{equation}
A_{min}(\nu,h)=F(\tau,{\bf x};\gamma).\label{Amin}
\end{equation}
This determines the contour lines $\nu=\nu(h)$ in Fig. \ref{fig:cf}.

\bibliography{sample}

\section*{Acknowledgements (not compulsory)}

This work was part of the project “WAMIZ – Waves in the Ice” funded by the Italian PNRA (grant PNRA18\_00109). This is also a contribution to the Year of Polar Prediction (YOPP), a flagship activity of the Polar Prediction Project (PPP), initiated by the World Weather Research Programme (WWRP) of the World Meteorological Organisation (WMO). We acknowledge the WMO WWRP for its role in coordinating this international research activity. We also acknowledge the activity carried out in the FP7 EU project ICE-ARC (grant agreement 603887).

The Project was carried out using CSK® Products © ASI (Italian Space Agency), delivered under the ASI licence to use ID 714 in the framework of COSMO-SkyMed Open Call for Science. The Sentinel-1A used in this work was freely delivered by ESA through the Copernicus Open Access Hub (\url{https://scihub.copernicus.eu/dhus/#/home}).

\section*{Author contributions statement}


\section*{Additional information}


The corresponding author is responsible for submitting a \href{http://www.nature.com/srep/policies/index.html#competing}{competing interests statement} on behalf of all authors of the paper. This statement must be included in the submitted article file.

%
%
%

\end{document}